\documentclass[sigconf]{acmart}
\usepackage[ruled,linesnumbered,resetcount]{algorithm2e}
\usepackage{tcolorbox}
\usepackage{algpseudocode}
\usepackage[utf8]{inputenc}
\usepackage{amsthm}
\usepackage{amsmath}
\usepackage{caption}
\usepackage{subcaption}

\usepackage{multicol}%
\usepackage{multirow}

\AtBeginDocument{%
  \providecommand\BibTeX{{%
    \normalfont B\kern-0.5em{\scshape i\kern-0.25em b}\kern-0.8em\TeX}}}

\acmConference[ICSE 2022]{The 44th International Conference on Software Engineering}{May 21–29, 2022}{Pittsburgh, PA, USA}

\begin{document}

%%
%% The "title" command has an optional parameter,
%% allowing the author to define a "short title" to be used in page headers.
\title{Toward the Analysis of Graph Neural Networks}

%%
%% The "author" command and its associated commands are used to define
%% the authors and their affiliations.
%% Of note is the shared affiliation of the first two authors, and the
%% "authornote" and "authornotemark" commands
%% used to denote shared contribution to the research.
\author{Thanh-Dat Nguyen}
\authornote{Both authors contributed equally to this research.}
\email{thanhdatn@student.unimelb.edu.au}
\affiliation{%
  \institution{University of Melbourne}
  \city{Melbourne}
  \state{Victoria}
  \country{Australia}
}
\author{Thanh Le-Cong}
\authornotemark[1]
\email{thanh.ld164834@sis.hust.edu.vn}
\affiliation{%
  \institution{Hanoi University of Science and Technology}
  \city{Hanoi}
  \country{Vietnam}
}

\author{ThanhVu H. Nguyen}
\email{tvn@gmu.edu}
\affiliation{%
  \institution{George Mason University}
  \city{Fairfax}
  \state{Virginia}
  \country{Vietnam}
}

\author{Xuan-Bach D. Le}
\email{bach.le@unimelb.edu.au}
\affiliation{%
  \institution{University of Melbourne}
  \city{Melbourne}
  \state{Victoria}
  \country{Australia}
}

\author{Quyet-Thang Huynh}
\email{thanghq@soict.hust.edu.vn}
\affiliation{%
  \institution{Hanoi University of Science and Technology}
  \city{Hanoi}
  \country{Vietnam}
}

%%
%% By default, the full list of authors will be used in the page
%% headers. Often, this list is too long, and will overlap
%% other information printed in the page headers. This command allows
%% the author to define a more concise list
%% of authors' names for this purpose.

%%
%% The abstract is a short summary of the work to be presented in the
%% article.
\begin{abstract}
% Graph Neural Networks (GNNs) have recently emerged as a powerful framework for graph-structured data and has been applied to many problems such as knowledge graph analysis, social networks recommendation, and even Covid19 detection and vaccine developments.
% However, unlike other types of neural networks such as Feed Forward Neural Networks (FFNNs), few analyses such as verification and property inferences exist, potentially due to dynamic behaviors of GNNs, which can take arbitrary graphs as input, whereas FFNNs only take fixed size numerical vectors as inputs. 

% In this paper, we propose an approach to analyze GNNs by converting them into FFNNs and reusing existing FFNNs analyses. We discuss various designs to ensure the scalability and accuracy of the conversion. 
% s.  We illustrate our method on a study case of node-classfication. We believe that our approach opens new research directions for understanding and analyzing GNNs. 
Graph Neural Networks (GNNs) have recently emerged as a robust framework for graph-structured data. They have been applied to many problems such as knowledge graph analysis, social networks recommendation, and even Covid19 detection and vaccine developments.
However, unlike other deep neural networks such as Feed Forward Neural Networks (FFNNs), few analyses such as verification and property inferences exist, potentially due to dynamic behaviors of GNNs, which can take arbitrary graphs as input, whereas FFNNs which only take fixed size numerical vectors as inputs. 

This paper proposes an approach to analyze GNNs by converting them into FFNNs and reusing existing FFNNs analyses. We discuss various designs to ensure the scalability and accuracy of the conversions.  We illustrate our method on a study case of node classification. We believe that our approach opens new research directions for understanding and analyzing GNNs.

%However, GNNs suffer from a lack of explainability for their prediction, which impedes applications of GNNs in safety-critical domains such as autonomous driving, banking, or medicine. This motivates a novel technique for providing human-intelligible explanations for GNNs.
%In this paper, we present a novel form of  GNN input properties, which consists of 3 components: a \textit{structure predicate} to constrict the structural space, an accompanying \textit{feature predicate} constraining the node feature space, and finally, the \textit{property} which the two prior predicates imply.
%By capturing GNN behavior as formal constraints, these input properties can be used to provide reinforce security condition of using trained models.
%Accompanying this novel form of input properties, we propose a method to effectively extract the two aforementioned predicate given a property by leveraging \textit{influential sub-structures} which highly affect model's prediction and by \textit{unrolling} GNNs to an equivalent Feed-forward neural network (FFNN).
%Finally, we illustrate our method on a study case of node-classfication.
% It can be seen that these input properties can serve as explanations for GNN's prediction. In particular, we propose to extract \textit{influential sub-structures}, which captures a significant part of the logic of GNNs, as . 
% Our approach then reduce GNNs to FFNNs and, followed by finding accompanying \textit{feature predicate} which implies certain output property. 

\end{abstract}

%%
%% The code below is generated by the tool at http://dl.acm.org/ccs.cfm.
%% Please copy and paste the code instead of the example below.
%%
\begin{CCSXML}
<ccs2012>
   <concept>
       <concept_id>10011007.10011074.10011099.10011102.10011103</concept_id>
       <concept_desc>Software and its engineering~Software testing and debugging</concept_desc>
       <concept_significance>500</concept_significance>
       </concept>
 </ccs2012>
\end{CCSXML}

% \ccsdesc[500]{Software and its engineering~Software testing and debugging}

%%
%% Keywords. The author(s) should pick words that accurately describe
%% the work being presented. Separate the keywords with commas.
\keywords{Property Inference, Formal Explanations, Graph Neural Networks}

\maketitle

\section{Introduction}
Deep Neural Networks (DNNs) have emerged as one of the most effective and modern approaches in solving problems, from common ones such as movies recommendations, image recognition to important ones such as collision control and "fake" news and information detection.  Just like software, these DNN models can be misused and attacked (e.g., small perturbations to the inputs can result in misclassifying results).  Thus, over the last decade, researchers have developed many powerful techniques to analyze DNNs, e.g.,  verifying that for certain inputs, a DNN will result in a certain output or have a desirable property, and more recently, inferring properties or facts to help explain behaviors of a DNN, which is typically treated as blackbox.

Despite the proliferation of DNNs analyses, most effective ones focus only on certain types of DNNS, such as Feedforward Neural Network (FFNNs), in which the inputs are presented as fixed-size vectors of numbers and a fixed structure network.  One of the more complicated DNNs that has recently been used in practice is Graph Neural Networks (GNNS), which take inputs as \emph{graphs} of various sizes (even each of the nodes in the graph is attached with information encoded as a vector of numbers) and have a dynamic network that depends on the structure and information from the input graphs. GNNs have been applied to solve many practical problems, e.g., knowledge graphs analysis~\cite{Vandenhende2020}, recommendation system for social networks~\cite{ying2018graph}, chemical and protein classification~\cite{Ranjan2019, Gilmer2017}, reasoning the structure of graphics and images~\cite{wang2018zero},  and even advanced COVID19 detection~\cite{zhu2021high, saha2021graphcovidnet} and vaccine development~\cite{cheung2020graph, hsieh2020drug, zhou2020artificial}.

Just as with standard DNNs, complex GNNs are often used as blackbox and can be vulnerable to adversary attacks, all of which lead to concerns about safety, fairness, and privacy of GNNs~\cite{zhu2019robust, ijcai2019-669, dai2018adversarial}, e.g., a new Covid vaccine developed by unknown and attacked-prone ML technique can only further increase doubts and hesitancy from the public. However, unlike popular FFNNs, in which there are many effective formal analyses, to the best of our knowledge, few exist for GNNs, potentially due to the  vast differences between the two types of networks.%\tvn{need additional evidence here,  is it true that very few analyses of GNNs?}.  
%\dat{First work of GNNExplanation has only been started as of 2019 by \cite{Ying2019}, few more works in 2020 of similar caliber \cite{luo2020parameterized, Vu2020a}, while neural attribution of FFNN has been introduced with \cite{Ribeiro2016} and earlier. Furthermore, to the best of our knowledge, no techniques for GNN verification or inferring property in form of formal mathematical formula has been introduced, while \cite{katz2017reluplex} for FFNN has been introduced in 2017}
%\dat{just reworded: analyses has been done on GNN, but not in form of tools, more like mathematical proof of reduced version of GNN}

In this paper, we propose an approach to analyze GNNs, both in verification and property inference, by converting GNNs to FFNNs and reusing developed techniques for FFNNs. While analyses explicitly designed for GNNs can be more efficient, they can be difficult and time-consuming to develop due to the differences between two types of networks. Thus, we believe our approach of leveraging existing efficient techniques and tools can be achieved quicker and also as effectively, as we can rely on existing powerful FFNN tools. This kind of approach is similar to many techniques in software engineering that encode the analysis task as a logical formula that can be efficiently analyzed by existing constraint solving techniques and tools (e.g., SAT and SMT solvers). 
\begin{figure}[t]
    \centering
    \includegraphics[width=0.8\linewidth]{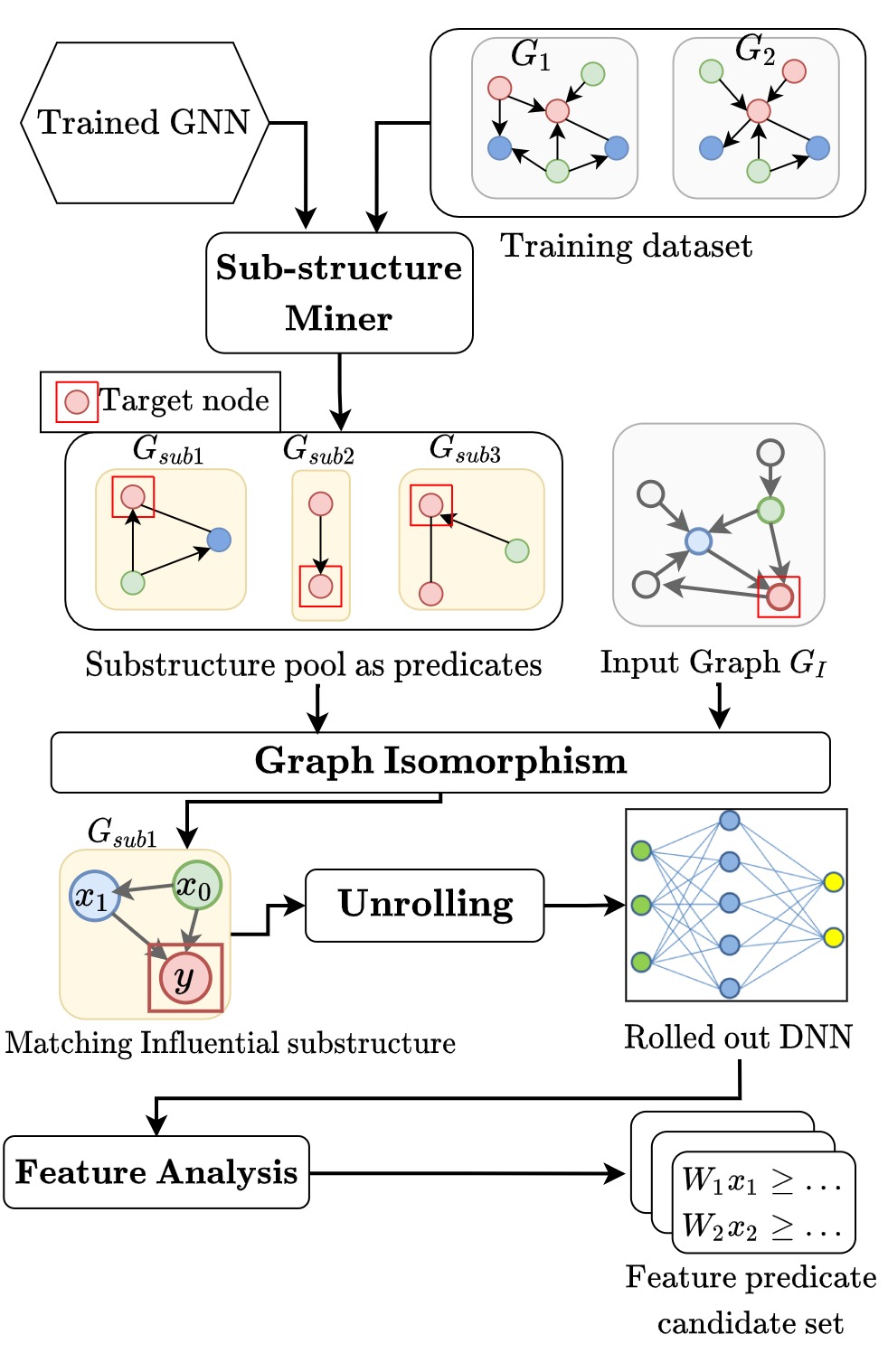}
    \caption{Overview}%, which consists of 5 main blocks: sub-structure miner, graph isomorphism predicate, unrolling, feature analysis and verification.}
    \label{fig:overview}
\end{figure}

Fig.~\ref{fig:overview} gives an overview of our idea.  The main challenge in analyzing GNNs and converting them to FFNNs is that the input graphs of a GNN can have various topological structures and the GNN has a dynamic structure depending on its input graphs.  
To solve this challenge, we mine influential sub-structures of input graphs to summarize the structural input space of a GNN.  
Then for each sub-structure, which represents a class of input graphs, we "unroll" the structure to create an equivalent FFNN for each node update operation of GNN, and then combine these FFNNs into a final FFNN representing the original GNN computation of the substructure. 
While influential substructure should contribute significantly toward GNN prediction, additional nodes that are non-influential can still affect the final computation result of the GNN. 
To deal with this, we find conditions on which the rolled out FFNN on the substructure and the original GNN is equivalent to ensure our results.

Finally, given the equivalent condition, we can now extend the existing DNN analyses to the rolled out FFNN of each substructure and obtain results for the original GNN.

\section{Technical Approach} 

% shown in Figure~\ref{fig:overview} using an example given in Fig.~\ref{fig:example}.

Existing verification techniques for DNNs attempt to check that the DNN satisfies a user-supplied property (e.g., a certain range over inputs results in a certain output). In contrast, property inference attempts to automatically infer such properties from the DNN.  In both cases, the property to be verified or inferred has the form $\texttt{pre} \implies \texttt{post}$, where \texttt{pre} is a condition over the inputs and \texttt{post} is certain requirement on the outputs.  Typically, we are interested in verifying or inferring the \texttt{pre} for some specific \texttt{post},  e.g., we want to find input conditions that make the DNN classifies an image as a "dog".

\begin{figure}
    \centering
    \includegraphics[width=\linewidth]{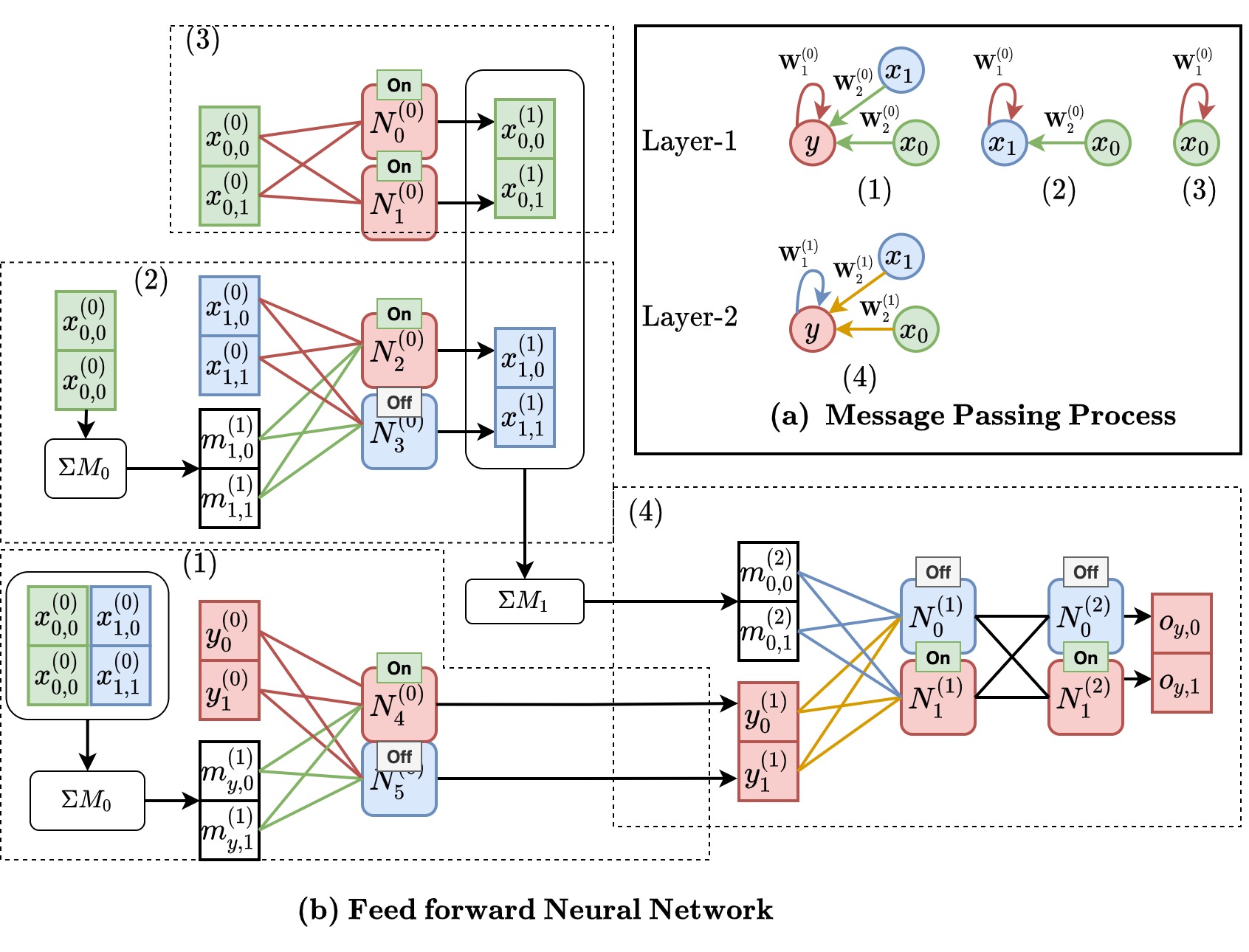}
    \caption{GNN message passing and unrolling}
    \label{fig:example}
\end{figure}

For this work, we will consider GNN models for the standard problem of graph node classification, which takes as input a graph $G$ and gives a classification $c$ for each node $v\in G$. %Here, assume that we are interested the classification of some \emph{target} node $y$. 
For such GNNs, the \texttt{pre} are \emph{input properties}, which are logical \emph{predicates} capturing common structure and features\footnote{Node features are attributes of nodes, e.g., if we take a node "professor" in academic graph, its attributes may be "name", "citations", "affiliation", and encoded as a numerical vector such as $\{0.1, 0.3, 0.4, 0.5\}$.} of the input graphs that lead to a certain classification of the target node.
Below we use a concrete example given in Fig.~\ref{fig:example} to describe the steps of our approach, whose overview is given in Fig.~\ref{fig:overview}.  
%Figure~\ref{fig:overview} gives an overview of our approach of converting a GNN into a FFNN to infer GNN input properties for the target node. 
%Figure~\ref{fig:example}a shows an example input graph with a red-colored target node.
%Within this setting, we can take arbitrary node $y \in V$ to be the node of interest for analysis without loss of generality, which we call target node (The red-colored node in Figure \ref{fig:overview}), \ref{fig:example}a, \ref{fig:example}b). 

\subsection{Substructure Mining}\label{subsec:substruct_mine}

Unlike an FFNN, a GNN does not have a fixed structure: it can take arbitrary graphs and the behaviors of GNN itself also change depending on the structure of the inputs (e.g., the influence of a node depends on its neighbors).  Thus, a direct, naïve way of converting a GNN to an FFNN does not scale as it would result in a different FFNN for each different input graph and the FFNN can also be very large if the input graph is large.

To solve this challenge, we will create FFNNs that support \emph{classes} of input graphs. 
We leverage existing works in network graphs and GNNs to mine common and influential \emph{substructures} from sample input graphs. 
These substructures are compact and summarize important behaviors of a GNN\cite{Ying2019, luo2020parameterized}, and existing works such as GNNExplainer and PGExplainer \cite{Ying2019, luo2020parameterized} can extract substructures that are important to each GNN prediction (e.g., sets of edges, nodes, and features that are important to the output of target node's prediction).
These substructures are crucial for the generation and analysis of FFNNs in subsequent tasks.
For example, comparing to individual graphs these influential substructures are compact, which are crucial to achieving FFNNs with manageable sizes.

Fig.~\ref{fig:overview} illustrates how we mine influential substructures (\textit{sub-structure miner}). For a trained GNN model, we take in a set of input graphs that have the desired classification and use an existing tool such as GNN explainer to extract subgraphs that are common from the input graphs and influential to the classification result of the target node. 
In Figure \ref{fig:overview}, from the two input graphs $G_1$ and $G_2$, GNNExplainer would extract three substructures  $G_{sub1}, G_{sub2}, G_{sub3}$ that have been determined to contribute significantly (influential) towards the target red-color node prediction while also being present in both graphs from the training dataset (frequented).

Our approach thus focuses on analyzing GNNs over input graphs that we have knowledge about (through the extracted influential substructures).
The better the sample of graph inputs we have, the larger and more accurate set of influential substructures we learn---this leads to larger classes of supported graphs.
Just as with most DNNs, sources of obtaining training samples vary, e.g., public benchmarks.  
If available, we can also reuse input graphs that were used to train the GNN models we are analyzing.

% \tvn{TODO: remember to discuss these training datasets somewhere, e.g., how do we get these training dataset,  how much is enough,  and how many substructures do we think it will generate ?  hundreds, thousands ??}
% \dat{These training dataset can be obtained by the GNN training process itself. Typical FSM mining would require thousands of graph. However, in reality it will be depending on each domain: more complicated domain where graph structure is more varied will need more samples in order for us to capture enough effective influential substructures.
% }

\subsection{Structure Predicates and Graph Isomorphism}\label{subsec:struct_predicate}
Many existing DNNs and program analyses encode problems into logical formulae that can be reasoned about using constraint solving.
Here, we encode the obtained graph substructures as logical predicates $\sigma_{struct}$, so that we can leverage existing automating reasoning tools such as SAT and SMT solvers.
An important use for these \emph{structure predicates} is to check if an input graph contains the considered substructures.
If it does, we are confident that our approach and result will hold; and if it does not, we can support it by adding it to our training data to learn about its  influential substructures.
These predicates are also a crucial part of the inferred properties that help explain the behaviors of the GNN to the user.

To determine if an input graph satisfies a substructure, we check if the graph and the substructure, which is also a graph, is \emph{isomorphic}.
By adapting existing work such as CFL-match~\cite{Bi2016a}, we can apply logical reasoning over the obtained structure predicates to check if there exist a mapping from the querying substructure graph to some subgraph of input graph that would make both graphs isomorphic.

Fig.~\ref{fig:overview} shows these steps.  For each obtained substructure, we create a predicate capturing that structure by using standard graph isomorphic checking. As an example structure predicate ensuring present of $G_{sub1}$ which has three nodes $x_0$ (green), $x_1$ (blue) and $y$ (red) to be matched to input graph $G = (V, E)$ where $V$ is the set of nodes and $E$ is the set of edges, can has the following form (note that we have omitted node-label checking for readability):
\begin{equation}
    \begin{split}
    \sigma_{struct, sub1}(V, E) = \exists  x_0, x_1, y \in V: \\
    \{(x_0, x_1), (x_0, x_1), (x_0, y)\} \subseteq E
    \end{split}
    \label{eq:concrete_struc}
\end{equation}

Concerning the implementation side, this can be done with existing analysis tools by transforming from a graph problem to âa satisfiability problem (e.g., nodes represented as boolean variables and edges as logical connections among variables).
Notice if we perform graph isomorphism checking on some input graph such as $G_1$ in the Figure ~\ref{fig:overview}, we will see that it is isomorphic to the  predicate of substructure $G_{sub1}$ because they share the same substructure. 

Obviously, all trained input sets would be isomorphic to at least one of the structure predicates.

\subsection{GNN Unrolling}\label{subsec:unrolling}

After obtaining influential substructures and their corresponding substructural predicates, we are now ready to create an FFNN to represent the GNN model.
As it turns out,  it is actually straightforward to convert a GNN with a fixed substructure directly to an FFNN.
We assume our GNN uses the the popular \emph{message passing process}\cite{Gilmer2017} adopted in most types of GNNs.
This process works by updating the value of a node in the graph based on the information of its neighboring nodes.
Then, to create an FFNN from a GNN with a set of substructures, we essentially create a FFNN to simulate how message passing is done on a substructure using the ``unroll'' technique similar to one introduced in~\cite{Jacoby2020} for RNN unrolling.
%In particular, each layer of GNN can be converted to an equivalent feedforward layer, in which features from source nodes are layer input and updated destination node features are the output.
Finally, we combine all FFNNs to obtain a final FFNN representing the original GNN (that supports graph inputs isomorphic to the considered substructures).

Figure~\ref{fig:example} shows how message passing works on the substructure $G_{sub1}$ obtained in Fig.~\ref{fig:overview}. 
Again, $G_{sub1}$ consists of 3 nodes $x_0, x_1$ and $y$,
for illustration purposes, we use a GNN with 2-layer and the weight $W$ to represent the values of features of each node.
For layer 1, the GNN contains three  message passing processes labelled (1), (2), (3) that correspond to the three nodes $y$, $x_1$ , and $x_0$.
The results of these message passing processes are updated as newly computed node features of $y, x_1$ and $x_0$, which are used for next layer.
For final layer 2, we only need to consider target node $y$'s message passing from the result of (1), (2) (3), followed by a simple linear transformation and we have a message processing process labeled (4) in Figure~\ref{fig:example}.

Now, we unroll each message passing process in layer $i$ of the GNN into a corresponding $i$-layer FFNN. 
For the three message-passing processes in Layer 1 in Figure~\ref{fig:example}a, we obtain the three 1-layer FFNNs shown in Figure~\ref{fig:example}b and for the message passing process in layer 2 in Figure~\ref{fig:example}a, we have the 2-layer FFNN shown in Figure~\ref{fig:example}b (4).  Finally, we connect these individual FFNNs to construct a final (large) FFNN as shown in Figure~\ref{fig:example}b to represent the original GNN.

%\tvn{Intuitively what does this resulting FFFN means with respect to the original GNN and the considered substructures?  If the already GNN classifies y as c, then this new FFNN does what? If you apply verification or property inference techniques to this FFNN, then what do these results mean for the original GNN? }
%\dat{This FFNN captures the computation of influential substructures  GNN }

\paragraph{Using Existing FFNN analyses}
We can apply existing analyses for FFNNs  to our rolled out FFNN.
For instance, we can apply the Prophecy tool~\cite{Gopinath2019} to infer properties for FFNNs.
This work derives predicates over the inputs of an FFNN, which convex regions over inputs values, that map to a desired output classification.
We can also apply FFNN verification tools such as Marabou (the successor of the popular Reluplex work~\cite{katz2017reluplex}) to check if an inferred or user-supplied property is correct.

For the running example in Figure~\ref{fig:example}, given some specific weights in Figure ~\ref{fig:example}a, running Prophecy on the resulting FFNN can gives the following predicates representing a convex region over the inputs space that result in the desired classification of the target node in the GNN.  Here the input $x_{i,j}$ of the FFNN represents the feature $j$ of node $x_i$ of the GNN. 
\begin{equation}
    \begin{split}
    \sigma_{inps} & = (x_{0,0} + x_{1,0} - x_{0,1} - x_{1,1} > 0) \\
    \wedge &  (x_{0,0} + x_{1,0} - 2x_{0,1} - 2x_{1,1} \leq 0) \\
    \wedge & (x_{0,0} -  x_{0,1} > 0) 
    \wedge (x_{0,0} - 2 x_{0,1} > 0) \\
    \wedge & (x_{0,0} + x_{1,0} + x_{2,0} - x_{0,1} - x_{1,1} - x_{2,1} > 0) \\
    \wedge & (x_{0,1} + x_{1,1} + x_{2,1} - 2x_{0,0} - 2x_{1,0} - 2x_{2,0} \leq 0)
\end{split}
\label{eq:concrete_infeat}
\end{equation}

%Note that the obtained results will be for the new FFNN, and we thus would need to convert them back to to our original GNN.

%We stress that:
%\begin{itemize}
%    \item The complete transformation from GNN to a corresponding FFNN can only be done when we have structural clues (i.e. we will not know the exact pattern of neurons $N_0^{(0)}, N_1^{(0)}, \ldots$ in Figure ~\ref{fig:example}b without a concrete structure).
%    \item The roll out on Figure ~\ref{fig:example}b, is not yet equivalent to the full graph given in Figure ~\ref{fig:overview}, so this is yet the full input predicate of GNN and still depends on extendable structural parts (i.e. nodes other than $x_0, x_1, y$ can be added to graph while still containing substructure $G_{sub1}$). We shows our extension in dealing with this problem in Subsection ~\ref{subsec:feat_analysis}
%\end{itemize}

% \tvn{What does this property mean with respect to the original graph G with target node y?} \thanh{This property can serve as a formal explanation to understand to understand why GNN makes a certain prediction for target node $y$ .Strutural property $\sigma_{struc}$ is a small subset of node and edges that are most influential for GNN's prediction(s), while feature property $\sigma_{feat}$ display a convex region in feature spaces, which implies a certain output prediction for target nodes $y$.} \thanh{I think applications of our inferred properties are similiar to PROPHECY (Corina's works) \cite{Gopinath2019}. But if we need to highlight these applications, we can add a section of "Applications"}

\subsection{Equivalent Analysis}\label{subsec:feat_analysis}

Ideally, the mined influential substructures truly represent the behaviors of the considered GNN and the obtained FFNN is thus equivalent to the GNN.
In practice, this does not happen as many nodes, especially those that are not part of the influential substructures but are neighbors with those in the substructure, can influence the final GNN classification result. Thus, we want to analyze how these  neighboring nodes can directly affect those in the influential substructures and thus the final result. 

Our experiences with GNNs show that a node in a influential substructure is affected by their "outside" neighbors (those that are not in the substructures) in two ways: the number of outside neighbors it has comparing to its total number of neighbors (\textbf{connectivity ratio}) and the \textbf{mean contribution} of the outside neighbors. 

Given this knowledge, we compute additional conditions over substructures to make them represent the GNN more accurately. To do this, we  use decision trees to compute predicates over the two features representing connectivity and mean contribution. We split the input graphs (e.g., used in the beginning to obtain substructures) into those that are and are not isomorphic to the substructures. With respect to each influential substructure, we collect the supporting input graph set from the training dataset. 
Following this, for each input in the supporting graph set, we perform two predictions of target node $y$'s output: 1) using only the influential substructure and 2) using the full input graph.
We collect statistics on \textbf{connectivity ratio} and \textbf{mean contribution} to predict whether the output on target node $y$ remains the same throughout two scenario.

Using this set of training data, we can leverage decision tree to determine conditions over the two features representing connectivity ratio and mean contribution that lead to equivalent or non-equivalent classification. Each paths in the tree represents an additional predicate that can help strengthen the substructures, allowing them to represent the GNN more accurately. 

\begin{figure}
    \centering
     \includegraphics[width=0.88\linewidth]{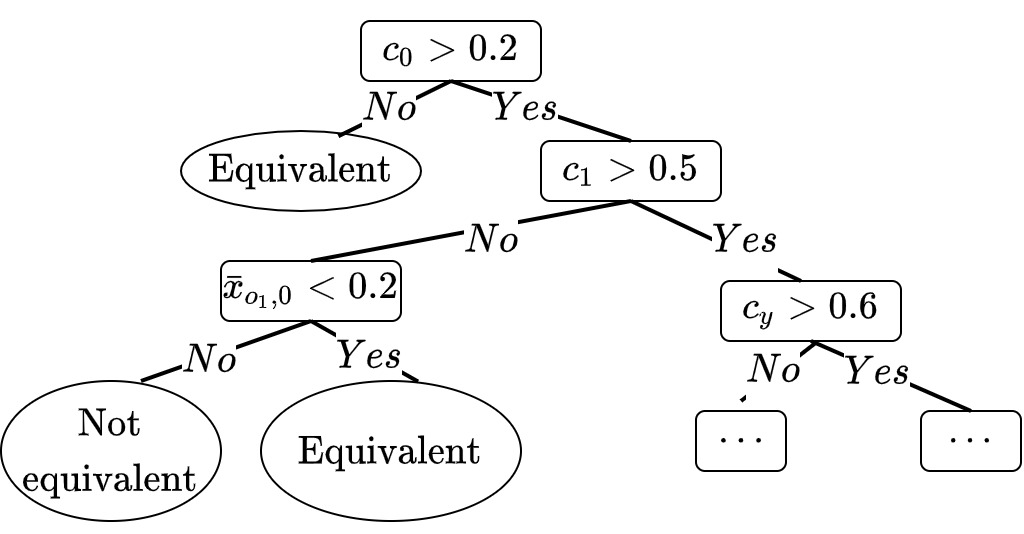}
    \caption{Example of using decision tree to predict whether substructure computation will be equivalent to the input graphs}
    \label{fig:decision_tree}
\end{figure}

For example, the decision tree in Fig.~\ref{fig:decision_tree} produces several predicates such as 
\begin{equation} \label{eq:feat_equiv}
\sigma_{\textit{feat\_equiv}} := c_0 > 0.2 \wedge c_1 \leq 0.5 \wedge \bar{x}_{o_1, 0} < 0.2, 
\end{equation}
\noindent  which says that node 0 with connectivity ratio $> 0.2$ , node 1 with connectivity ratio $<= 0.5$, and the mean contribution of feature 0 in node 0 $\ge 0.2$ are likely needed as conjunction for the feature predicate on the substructure holds for all graphs.
Thus, this approach allows us to obtain a set $\sigma_{inps}$ of predicates to strengthen the substructure predicates, ensuring they represent the GNN more accurately.

\subsection{GNN Property}
At the end, our approach produces a \textit{property} $\sigma$ of a given GNN in form of 
\begin{equation*}
    \sigma_{struct} \sigma \wedge \sigma_{inps} \wedge \sigma_{feat\_equiv} \implies Q,
\end{equation*}

\noindent where $\sigma_{struct}$ is predicate capturing graph isomorphism (Section~\ref{subsec:struct_predicate}), $\sigma_{inp}$ are input properties of the converted FFNN (Section~\ref{subsec:unrolling}), $\sigma_{feat\_equiv}$ is the additional constraints helping the FFNN more accurate to the original GNN (Section~\ref{subsec:feat_analysis}), and Q is the output property of some target node $y$ (e.g. $o_{y,1} < o_{y,2}$).
This means for an an input graph with target node $y$ that is isomorphic to some of the mined substructure (satisfies $\sigma_{struct})$, has certain requirements about neighboring substructure nodes (satisfies $\sigma_{feat\_equiv}$),  with node features lie within certain regions (satisfies $\sigma_{inps}$), then this graph will have the property $Q$ on its target node.

%\subsection{Verification}
%We note that while $\sigma_{feat\_dnn}$ is completely verifiable by existing FFNN verification tools such as Reluplex \cite{katz2017reluplex} or Marabou\cite{katz2019marabou}.
%However, in the computation process of GNN, there exists quadratic terms, which can not be formally verified with existing tools (see Figure ~\ref{fig:unroll_small}, where $m_{y,0}^{(1)}$ consists of a quadratic term of 2 variables - the contribution factor $c_y = \frac{1}{n}$ and $x_{1,0}$ node feature).
%While we circumvented verifying equivalent feature predicate by leveraging statistical tools (i.e. decision tree), we believe that formally verifying non-linear terms is a desirable feature of DNN verification tools that need to be developed. Which we list as one of our future directions.\tvn{unclear, the decision tree is used to guess some predicates in feat\_equiv,  it's not related to feat\_dnn.  So why are they discussed together?}

\section{Future Plan}
Currently, we only have worked out our idea on several small examples by hand. We are now moving to implementing these ideas to automate the process.  After that, we will evaluate the approach with existing GNN benchmarks.

We anticipate several challenges that would arise in this direction. First, for complex GNNs, the converted FFNNs might be too complex and contain non-trivial, e.g., nonlinear-arithmetic.  These would give difficulties to standard FFNN verification tools such as Reluplex. 
Second, we use sample inputs to mine substructures and feature predicates, and thus can obtain inaccurate results. While we might be able to obtain "groundtruths" or manually check results of small GNNs, we will not have an effective way to formally verify our results on complex  and real-world GNNs. Third, obtaining realistic benchmarks for GNNs might be more difficult as they are not as abundent and well-studied as benchmarks of FFNNs. However, we can start with existing dataset from the literature such as those from~\cite{wang2018zero} for autnomous driving and \cite{Gilmer2017, zitnik2018modeling} for drug interactions.

Finally, there is always a chance that this entire approach does not work well in practice, e.g., it does not scale or becomes too inaccurate for converting complex GNNs. It might be that designing algorithms directly to solve GNN would give more benefits in the long run. However, as with any research problem, especially those with few existing attempts, we have to start somewhere, and converting it to something we do know how to do well seems to be a good place to start.  

\bibliographystyle{ACM-Reference-Format}
\bibliography{sample-authordraft}

%%% -*-BibTeX-*-
%%% Do NOT edit. File created by BibTeX with style
%%% ACM-Reference-Format-Journals [18-Jan-2012].

\begin{thebibliography}{20}

%%% ====================================================================
%%% NOTE TO THE USER: you can override these defaults by providing
%%% customized versions of any of these macros before the \bibliography
%%% command.  Each of them MUST provide its own final punctuation,
%%% except for \shownote{}, \showDOI{}, and \showURL{}.  The latter two
%%% do not use final punctuation, in order to avoid confusing it with
%%% the Web address.
%%%
%%% To suppress output of a particular field, define its macro to expand
%%% to an empty string, or better, \unskip, like this:
%%%
%%% \newcommand{\showDOI}[1]{\unskip}   % LaTeX syntax
%%%
%%% \def \showDOI #1{\unskip}           % plain TeX syntax
%%%
%%% ====================================================================

\ifx \showCODEN    \undefined \def \showCODEN     #1{\unskip}     \fi
\ifx \showDOI      \undefined \def \showDOI       #1{#1}\fi
\ifx \showISBNx    \undefined \def \showISBNx     #1{\unskip}     \fi
\ifx \showISBNxiii \undefined \def \showISBNxiii  #1{\unskip}     \fi
\ifx \showISSN     \undefined \def \showISSN      #1{\unskip}     \fi
\ifx \showLCCN     \undefined \def \showLCCN      #1{\unskip}     \fi
\ifx \shownote     \undefined \def \shownote      #1{#1}          \fi
\ifx \showarticletitle \undefined \def \showarticletitle #1{#1}   \fi
\ifx \showURL      \undefined \def \showURL       {\relax}        \fi
% The following commands are used for tagged output and should be
% invisible to TeX
\providecommand\bibfield[2]{#2}
\providecommand\bibinfo[2]{#2}
\providecommand\natexlab[1]{#1}
\providecommand\showeprint[2][]{arXiv:#2}

\bibitem[\protect\citeauthoryear{Bi, Chang, Lin, Qin, and Zhang}{Bi
  et~al\mbox{.}}{2016}]%
        {Bi2016a}
\bibfield{author}{\bibinfo{person}{Fei Bi}, \bibinfo{person}{Lijun Chang},
  \bibinfo{person}{Xuemin Lin}, \bibinfo{person}{Lu Qin}, {and}
  \bibinfo{person}{Wenjie Zhang}.} \bibinfo{year}{2016}\natexlab{}.
\newblock \showarticletitle{{Efficient subgraph matching by postponing
  Cartesian products}}.
\newblock \bibinfo{journal}{\emph{Proceedings of the ACM SIGMOD International
  Conference on Management of Data}}  \bibinfo{volume}{26-June-20}
  (\bibinfo{year}{2016}), \bibinfo{pages}{1199--1214}.
\newblock
\showISSN{07308078}
\urldef\tempurl%
\url{https://doi.org/10.1145/2882903.2915236}
\showDOI{\tempurl}


\bibitem[\protect\citeauthoryear{Cheung and Moura}{Cheung and Moura}{2020}]%
        {cheung2020graph}
\bibfield{author}{\bibinfo{person}{Mark Cheung} {and}
  \bibinfo{person}{Jos{\'e}~MF Moura}.} \bibinfo{year}{2020}\natexlab{}.
\newblock \showarticletitle{Graph Neural Networks for COVID-19 Drug Discovery}.
  In \bibinfo{booktitle}{\emph{2020 IEEE International Conference on Big Data
  (Big Data)}}. IEEE, \bibinfo{pages}{5646--5648}.
\newblock


\bibitem[\protect\citeauthoryear{Dai, Li, Tian, Huang, Wang, Zhu, and Song}{Dai
  et~al\mbox{.}}{2018}]%
        {dai2018adversarial}
\bibfield{author}{\bibinfo{person}{Hanjun Dai}, \bibinfo{person}{Hui Li},
  \bibinfo{person}{Tian Tian}, \bibinfo{person}{Xin Huang},
  \bibinfo{person}{Lin Wang}, \bibinfo{person}{Jun Zhu}, {and}
  \bibinfo{person}{Le Song}.} \bibinfo{year}{2018}\natexlab{}.
\newblock \showarticletitle{Adversarial attack on graph structured data}. In
  \bibinfo{booktitle}{\emph{International conference on machine learning}}.
  PMLR, \bibinfo{pages}{1115--1124}.
\newblock


\bibitem[\protect\citeauthoryear{Gilmer, Schoenholz, Riley, Vinyals, and
  Dahl}{Gilmer et~al\mbox{.}}{2017}]%
        {Gilmer2017}
\bibfield{author}{\bibinfo{person}{Justin Gilmer}, \bibinfo{person}{Samuel~S.
  Schoenholz}, \bibinfo{person}{Patrick~F. Riley}, \bibinfo{person}{Oriol
  Vinyals}, {and} \bibinfo{person}{George~E. Dahl}.}
  \bibinfo{year}{2017}\natexlab{}.
\newblock \showarticletitle{{Neural Message Passing for Quantum Chemistry}}.
\newblock  (\bibinfo{date}{4} \bibinfo{year}{2017}).
\newblock
\urldef\tempurl%
\url{http://arxiv.org/abs/1704.01212}
\showURL{%
\tempurl}


\bibitem[\protect\citeauthoryear{Gopinath, Converse, Pasareanu, and
  Taly}{Gopinath et~al\mbox{.}}{2019}]%
        {Gopinath2019}
\bibfield{author}{\bibinfo{person}{Divya Gopinath}, \bibinfo{person}{Hayes
  Converse}, \bibinfo{person}{Corina~S. Pasareanu}, {and}
  \bibinfo{person}{Ankur Taly}.} \bibinfo{year}{2019}\natexlab{}.
\newblock \showarticletitle{{Property Inference For Deep Neural Networks}}.
\newblock \bibinfo{journal}{\emph{2019 34th IEEE/ACM International Conference
  on Automated Software Engineering (ASE)}} (\bibinfo{date}{4}
  \bibinfo{year}{2019}), \bibinfo{pages}{797--809}.
\newblock
\showISBNx{978-1-7281-2508-4}
\urldef\tempurl%
\url{https://doi.org/10.1109/ASE.2019.00079}
\showDOI{\tempurl}


\bibitem[\protect\citeauthoryear{Hsieh, Wang, Chen, Zhao, Savitz, Jiang, Tang,
  and Kim}{Hsieh et~al\mbox{.}}{2020}]%
        {hsieh2020drug}
\bibfield{author}{\bibinfo{person}{Kang-Lin Hsieh}, \bibinfo{person}{Yinyin
  Wang}, \bibinfo{person}{Luyao Chen}, \bibinfo{person}{Zhongming Zhao},
  \bibinfo{person}{Sean Savitz}, \bibinfo{person}{Xiaoqian Jiang},
  \bibinfo{person}{Jing Tang}, {and} \bibinfo{person}{Yejin Kim}.}
  \bibinfo{year}{2020}\natexlab{}.
\newblock \showarticletitle{Drug repurposing for covid-19 using graph neural
  network with genetic, mechanistic, and epidemiological validation}.
\newblock \bibinfo{journal}{\emph{Research Square}} (\bibinfo{year}{2020}).
\newblock


\bibitem[\protect\citeauthoryear{Jacoby, Barrett, and Katz}{Jacoby
  et~al\mbox{.}}{2020}]%
        {Jacoby2020}
\bibfield{author}{\bibinfo{person}{Yuval Jacoby}, \bibinfo{person}{Clark
  Barrett}, {and} \bibinfo{person}{Guy Katz}.} \bibinfo{year}{2020}\natexlab{}.
\newblock \showarticletitle{{Verifying Recurrent Neural Networks using
  Invariant Inference}}.
\newblock \bibinfo{journal}{\emph{Lecture Notes in Computer Science (including
  subseries Lecture Notes in Artificial Intelligence and Lecture Notes in
  Bioinformatics)}}  \bibinfo{volume}{12302 LNCS} (\bibinfo{date}{4}
  \bibinfo{year}{2020}), \bibinfo{pages}{57--74}.
\newblock
\showISBNx{9783030591519}
\showISSN{16113349}
\urldef\tempurl%
\url{https://doi.org/10.1007/978-3-030-59152-6{\_}3}
\showDOI{\tempurl}


\bibitem[\protect\citeauthoryear{Katz, Barrett, Dill, Julian, and
  Kochenderfer}{Katz et~al\mbox{.}}{2017}]%
        {katz2017reluplex}
\bibfield{author}{\bibinfo{person}{Guy Katz}, \bibinfo{person}{Clark Barrett},
  \bibinfo{person}{David~L Dill}, \bibinfo{person}{Kyle Julian}, {and}
  \bibinfo{person}{Mykel~J Kochenderfer}.} \bibinfo{year}{2017}\natexlab{}.
\newblock \showarticletitle{Reluplex: An efficient SMT solver for verifying
  deep neural networks}. In \bibinfo{booktitle}{\emph{International Conference
  on Computer Aided Verification}}. Springer, \bibinfo{pages}{97--117}.
\newblock


\bibitem[\protect\citeauthoryear{Luo, Cheng, Xu, Yu, Zong, Chen, and Zhang}{Luo
  et~al\mbox{.}}{2020}]%
        {luo2020parameterized}
\bibfield{author}{\bibinfo{person}{Dongsheng Luo}, \bibinfo{person}{Wei Cheng},
  \bibinfo{person}{Dongkuan Xu}, \bibinfo{person}{Wenchao Yu},
  \bibinfo{person}{Bo Zong}, \bibinfo{person}{Haifeng Chen}, {and}
  \bibinfo{person}{Xiang Zhang}.} \bibinfo{year}{2020}\natexlab{}.
\newblock \showarticletitle{Parameterized Explainer for Graph Neural Network}.
\newblock \bibinfo{journal}{\emph{Advances in Neural Information Processing
  Systems}}  \bibinfo{volume}{33} (\bibinfo{year}{2020}).
\newblock


\bibitem[\protect\citeauthoryear{Ranjan, Sanyal, and Talukdar}{Ranjan
  et~al\mbox{.}}{2019}]%
        {Ranjan2019}
\bibfield{author}{\bibinfo{person}{Ekagra Ranjan}, \bibinfo{person}{Soumya
  Sanyal}, {and} \bibinfo{person}{Partha~Pratim Talukdar}.}
  \bibinfo{year}{2019}\natexlab{}.
\newblock \showarticletitle{{ASAP: Adaptive Structure Aware Pooling for
  Learning Hierarchical Graph Representations}}.
\newblock  (\bibinfo{date}{11} \bibinfo{year}{2019}).
\newblock
\urldef\tempurl%
\url{http://arxiv.org/abs/1911.07979}
\showURL{%
\tempurl}


\bibitem[\protect\citeauthoryear{Saha, Mukherjee, Singh, Ahmadian, Ferrara, and
  Sarkar}{Saha et~al\mbox{.}}{2021}]%
        {saha2021graphcovidnet}
\bibfield{author}{\bibinfo{person}{Pritam Saha}, \bibinfo{person}{Debadyuti
  Mukherjee}, \bibinfo{person}{Pawan~Kumar Singh}, \bibinfo{person}{Ali
  Ahmadian}, \bibinfo{person}{Massimiliano Ferrara}, {and} \bibinfo{person}{Ram
  Sarkar}.} \bibinfo{year}{2021}\natexlab{}.
\newblock \showarticletitle{GraphCovidNet: A graph neural network based model
  for detecting COVID-19 from CT scans and X-rays of chest}.
\newblock \bibinfo{journal}{\emph{Scientific Reports}} \bibinfo{volume}{11},
  \bibinfo{number}{1} (\bibinfo{year}{2021}), \bibinfo{pages}{1--16}.
\newblock


\bibitem[\protect\citeauthoryear{Vandenhende, Georgoulis, Van~Gansbeke,
  Proesmans, Dai, and Van~Gool}{Vandenhende et~al\mbox{.}}{2020}]%
        {Vandenhende2020}
\bibfield{author}{\bibinfo{person}{Simon Vandenhende},
  \bibinfo{person}{Stamatios Georgoulis}, \bibinfo{person}{Wouter
  Van~Gansbeke}, \bibinfo{person}{Marc Proesmans}, \bibinfo{person}{Dengxin
  Dai}, {and} \bibinfo{person}{Luc Van~Gool}.} \bibinfo{year}{2020}\natexlab{}.
\newblock \showarticletitle{{Multi-Task Learning for Dense Prediction Tasks: A
  Survey}}.
\newblock  (\bibinfo{year}{2020}), \bibinfo{pages}{1--20}.
\newblock
\urldef\tempurl%
\url{http://arxiv.org/abs/2004.13379}
\showURL{%
\tempurl}


\bibitem[\protect\citeauthoryear{Wang, Ye, and Gupta}{Wang
  et~al\mbox{.}}{2018}]%
        {wang2018zero}
\bibfield{author}{\bibinfo{person}{Xiaolong Wang}, \bibinfo{person}{Yufei Ye},
  {and} \bibinfo{person}{Abhinav Gupta}.} \bibinfo{year}{2018}\natexlab{}.
\newblock \showarticletitle{Zero-shot recognition via semantic embeddings and
  knowledge graphs}. In \bibinfo{booktitle}{\emph{Proceedings of the IEEE
  conference on computer vision and pattern recognition}}.
  \bibinfo{pages}{6857--6866}.
\newblock


\bibitem[\protect\citeauthoryear{Wu, Wang, Tyshetskiy, Docherty, Lu, and
  Zhu}{Wu et~al\mbox{.}}{2019}]%
        {ijcai2019-669}
\bibfield{author}{\bibinfo{person}{Huijun Wu}, \bibinfo{person}{Chen Wang},
  \bibinfo{person}{Yuriy Tyshetskiy}, \bibinfo{person}{Andrew Docherty},
  \bibinfo{person}{Kai Lu}, {and} \bibinfo{person}{Liming Zhu}.}
  \bibinfo{year}{2019}\natexlab{}.
\newblock \showarticletitle{Adversarial Examples for Graph Data: Deep Insights
  into Attack and Defense}. In \bibinfo{booktitle}{\emph{Proceedings of the
  Twenty-Eighth International Joint Conference on Artificial Intelligence,
  {IJCAI-19}}}. \bibinfo{publisher}{International Joint Conferences on
  Artificial Intelligence Organization}, \bibinfo{pages}{4816--4823}.
\newblock
\urldef\tempurl%
\url{https://doi.org/10.24963/ijcai.2019/669}
\showDOI{\tempurl}


\bibitem[\protect\citeauthoryear{Ying, Bourgeois, You, Zitnik, and
  Leskovec}{Ying et~al\mbox{.}}{2019}]%
        {Ying2019}
\bibfield{author}{\bibinfo{person}{Rex Ying}, \bibinfo{person}{Dylan
  Bourgeois}, \bibinfo{person}{Jiaxuan You}, \bibinfo{person}{Marinka Zitnik},
  {and} \bibinfo{person}{Jure Leskovec}.} \bibinfo{year}{2019}\natexlab{}.
\newblock \showarticletitle{{GNNExplainer: Generating Explanations for Graph
  Neural Networks}}.
\newblock  (\bibinfo{date}{3} \bibinfo{year}{2019}).
\newblock
\urldef\tempurl%
\url{https://arxiv.org/abs/1903.03894}
\showURL{%
\tempurl}


\bibitem[\protect\citeauthoryear{Ying, He, Chen, Eksombatchai, Hamilton, and
  Leskovec}{Ying et~al\mbox{.}}{2018}]%
        {ying2018graph}
\bibfield{author}{\bibinfo{person}{Rex Ying}, \bibinfo{person}{Ruining He},
  \bibinfo{person}{Kaifeng Chen}, \bibinfo{person}{Pong Eksombatchai},
  \bibinfo{person}{William~L Hamilton}, {and} \bibinfo{person}{Jure Leskovec}.}
  \bibinfo{year}{2018}\natexlab{}.
\newblock \showarticletitle{Graph convolutional neural networks for web-scale
  recommender systems}. In \bibinfo{booktitle}{\emph{Proceedings of the 24th
  ACM SIGKDD International Conference on Knowledge Discovery \& Data Mining}}.
  \bibinfo{pages}{974--983}.
\newblock


\bibitem[\protect\citeauthoryear{Zhou, Wang, Tang, Nussinov, and Cheng}{Zhou
  et~al\mbox{.}}{2020}]%
        {zhou2020artificial}
\bibfield{author}{\bibinfo{person}{Yadi Zhou}, \bibinfo{person}{Fei Wang},
  \bibinfo{person}{Jian Tang}, \bibinfo{person}{Ruth Nussinov}, {and}
  \bibinfo{person}{Feixiong Cheng}.} \bibinfo{year}{2020}\natexlab{}.
\newblock \showarticletitle{Artificial intelligence in COVID-19 drug
  repurposing}.
\newblock \bibinfo{journal}{\emph{The Lancet Digital Health}}
  (\bibinfo{year}{2020}).
\newblock


\bibitem[\protect\citeauthoryear{Zhu, Zhang, Cui, and Zhu}{Zhu
  et~al\mbox{.}}{2019}]%
        {zhu2019robust}
\bibfield{author}{\bibinfo{person}{Dingyuan Zhu}, \bibinfo{person}{Ziwei
  Zhang}, \bibinfo{person}{Peng Cui}, {and} \bibinfo{person}{Wenwu Zhu}.}
  \bibinfo{year}{2019}\natexlab{}.
\newblock \showarticletitle{Robust graph convolutional networks against
  adversarial attacks}. In \bibinfo{booktitle}{\emph{Proceedings of the 25th
  ACM SIGKDD International Conference on Knowledge Discovery \& Data Mining}}.
  \bibinfo{pages}{1399--1407}.
\newblock


\bibitem[\protect\citeauthoryear{Zhu, Bukharin, Xie, Santillana, Yang, and
  Xie}{Zhu et~al\mbox{.}}{2021}]%
        {zhu2021high}
\bibfield{author}{\bibinfo{person}{Shixiang Zhu}, \bibinfo{person}{Alexander
  Bukharin}, \bibinfo{person}{Liyan Xie}, \bibinfo{person}{Mauricio
  Santillana}, \bibinfo{person}{Shihao Yang}, {and} \bibinfo{person}{Yao Xie}.}
  \bibinfo{year}{2021}\natexlab{}.
\newblock \showarticletitle{High-resolution Spatio-temporal Model for
  County-level COVID-19 Activity in the US}.
\newblock \bibinfo{journal}{\emph{ACM Transactions on Management Information
  Systems (TMIS)}} \bibinfo{volume}{12}, \bibinfo{number}{4}
  (\bibinfo{year}{2021}), \bibinfo{pages}{1--20}.
\newblock


\bibitem[\protect\citeauthoryear{Zitnik, Agrawal, and Leskovec}{Zitnik
  et~al\mbox{.}}{2018}]%
        {zitnik2018modeling}
\bibfield{author}{\bibinfo{person}{Marinka Zitnik}, \bibinfo{person}{Monica
  Agrawal}, {and} \bibinfo{person}{Jure Leskovec}.}
  \bibinfo{year}{2018}\natexlab{}.
\newblock \showarticletitle{Modeling polypharmacy side effects with graph
  convolutional networks}.
\newblock \bibinfo{journal}{\emph{Bioinformatics}} \bibinfo{volume}{34},
  \bibinfo{number}{13} (\bibinfo{year}{2018}), \bibinfo{pages}{i457--i466}.
\newblock


\end{thebibliography}
\end{document}